\documentclass[aps,twocolumn,superscriptaddress]{revtex4}



\usepackage{graphicx}
\usepackage{dcolumn}

\begin{document}

\title{Filtering of complex systems using overlapping tree networks}

\author{Antonios Garas}
\email[]{agara@physics.auth.gr}
\affiliation{Department of Physics, University of Thessaloniki, 54124 Thessaloniki Greece.}

\author{Panos Argyrakis}
\email[]{panos@physics.auth.gr}
\affiliation{Department of Physics, University of Thessaloniki, 54124 Thessaloniki Greece.}

\date{\today}

\begin{abstract}

We introduce a technique that is capable to filter out information from complex systems, by mapping them to networks, and extracting a subgraph with the strongest links. This idea is based on the Minimum Spanning Tree, and it can be applied to sets of graphs that have as links different sets of interactions among the system's elements, which are described as network nodes. It can also be applied to correlation-based graphs, where the links are weighted and represent the correlation strength between all pairs of nodes. We applied this method to the European scientific collaboration network, which is composed of all the projects supported by the European Framework Program FP6, and also to the correlation-based network of the 100 highest capitalized stocks traded in the NYSE. For both cases we identified meaningful structures, such as a strongly interconnected community of countries that play important role in the collaboration network, and clusters of stocks belonging to different sectors of economic activity, which gives significant information about the investigated systems.

\end{abstract}

\maketitle

\section{Introduction}

The study of complex systems, a name that is frequently used for systems having a large number of elements that interact in (usually) non trivial ways, has been greatly advanced in recent years by the use of graph theory~\cite{bb:West}. One can map a complex system to a complex network by representing the interacting elements of the system with nodes and their interactions with links between the nodes. Examples of complex systems that have been recently investigated in this perspective include the Internet~\cite{bb:Faloutsos,bb:ShaiCarmi}, the World Wide Web~\cite{bb:AlbertJeongBarabasi}, communication networks~\cite{bb:PastorVespignani}, food webs~\cite{bb:Garlaschelli}, sexual contacts among individuals~\cite{bb:Liljeros}, economic networks~\cite{book:MantegnaStanley,bb:Mantegna,bb:Bonanno,bb:Onnela,bb:Garas,bb:TumminelloChaos,bb:Tumminello,bb:GarasArgyrakisHavlin}, the network of collaborations in EU funded projects~\cite{bb:MendesFP5, bb:GarasFP}, etc. 

Complex systems are not always static, meaning that they may evolve dynamically over time. This evolution can provide a wealth of information about the processes driving the system. One way to study such systems is to record the time dependence of some specific and well defined property, and thus obtain a set of time series that are able to depict the time evolution of the entire system. These time series can be transformed into a graph by using a similarity measure based in the cross-correlation among its elements, as very effectively has been done, for example, for the study of equity portfolios~\cite{book:MantegnaStanley,bb:Mantegna,bb:Bonanno,bb:Onnela,bb:Garas,bb:TumminelloChaos}, which has led to an established method for the investigation of financial systems. 
Therefore, we can use the cross-correlations between variables to create a correlation based network from any complex system that we have data available. On the other hand, if the system is static, meaning that its interactions do not change with time, but there are different kinds of interactions among its elements, we can create a static graph for each different interaction, and study this set of graphs. 

In this work we describe an approach that can be used for both types of the aforementioned analyses. It is based on the Minimum Spanning Tree (MST) technique~\cite{bb:Mantegna}, but it allows the creation of a more representative network of the system (compared to a simple MST) that maintains information about its dynamics and its temporal evolution. As application, we use this method to analyze the European scientific collaboration network for the projects carried out with the support of the Framework Program FP6, and the network of the 100 highest capitalized stocks traded in the NYSE in the period 1995-2003.

\section{The Method}

Any complex system with $N$ interacting elements can be mapped to a network with $N$ nodes. The links connecting the nodes of the network represent the interactions among the system elements, and the strength of these interactions is used as weights of the links between the nodes. The total number of links of a network depends on the information that we have about interactions between its nodes. For the special case that we know the interaction strength between all pairs of nodes, as we do in correlation based graphs, then the network is fully connected and has $N(N-1)/2$ number of links. For such cases it is essential to use filtering techniques in order to reduce the number of connections, so that we can study properties of the network that generally are hidden due to its complexity. The most drastic filtering of a network can be achieved by the extraction of the MST~\cite{book:Papadimitriou}, a technique that has been used extensively in the literature for the study of financial correlation matrices~\cite{bb:Mantegna,bb:Bonanno,bb:Onnela,bb:Garas,bb:TumminelloChaos}, leading to the identification of clusters of stocks that result in a meaningful taxonomy. The MST was recently used to extract the mode of collaboration in research projects funded by the European Commission Framework Programs~\cite{bb:GarasFP}. The MST is a graph with the same number of nodes $N$ as the original network but having a total of only $N-1$ edges, with a total minimum weight. It is constructed starting with to the disconnected graph that contains all the nodes of the network, and then by adding links  in increasing weight order, as long as they do not form loops, until all nodes are connected. Such structure is a much simpler graph than the fully connected network, but it still gives us interesting information about the system.

Depending on the structure of the system under investigation, and on the available data, we could create a series of representative networks that contain information about the evolution and the dynamics of the original system. For the case of networks that are stationary, we can construct different snapshots of them by using as links different interactions among the. This allows us to calculate MSTs separately for all different snapshots of the system. On the other hand, for the case of correlation based graphs, where the network is constructed using time series data, we divide the time series into smaller segments (time windows) and construct different network snapshots of the system at different time periods. Let us name $E$ the set of all the links present in each calculated MST, and let us assume that we have calculated a total of $M$ MSTs from the $M$ different snapshots of the original network that we can obtain. We can create the graph $G$ that has as links the union of all sets $E$ for all the $M$ trees that we have calculated \[G=\bigcup_{\tau=1}^{M}E^{\tau}.\] The number of links of graph $G$ lies inside the interval $[N-1, M(N-1)]$, where $N-1$ is the number of links of one MST with $N$ nodes and $M(N-1)$ is the total number of links of $M$ different MSTs. Obviously, the graph $G$ will have $N-1$ links if the investigated system is so stable that for all the $M$ investigated snapshots we find the same MST, while it will have $M(N-1)$ links if the system is so unstable that for all the $M$ investigated snapshots we find completely different MSTs. For the case of a highly unstable system, if we choose $M\geq N/2$, we will get a fully connected graph with $N(N-1)/2$ links.

\begin{figure}
\resizebox{1\columnwidth}{!}{%
\includegraphics{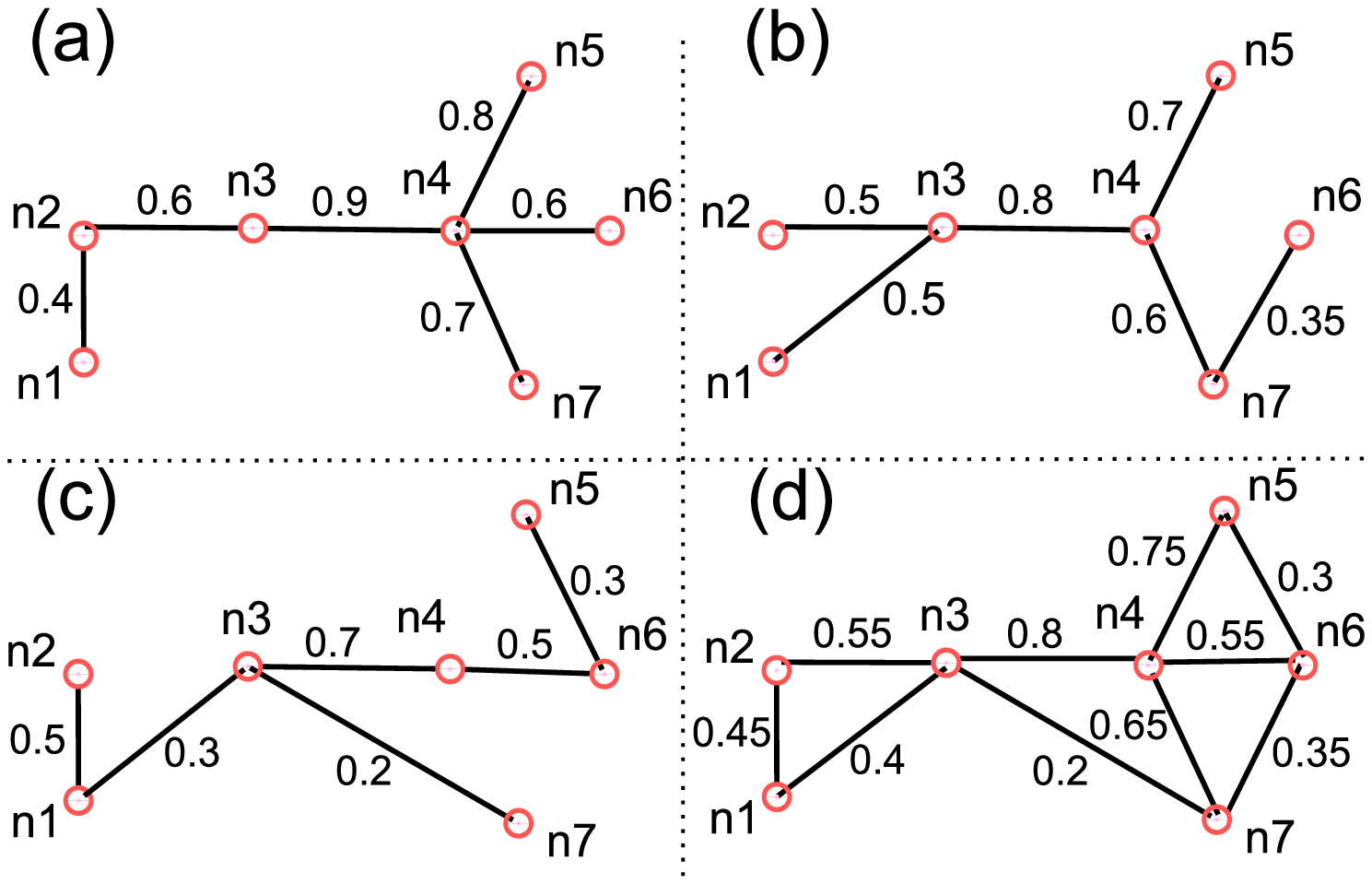}
}
\caption{An example of creation of the Overlapping Tree Network (OTN). Three trees - (a), (b), and (c) - are combined to form the Overlapping Tree Network (d).}
\label{Fig:1}
\end{figure}

\begin{figure*}
\resizebox{1.5\columnwidth}{!}{%
\includegraphics{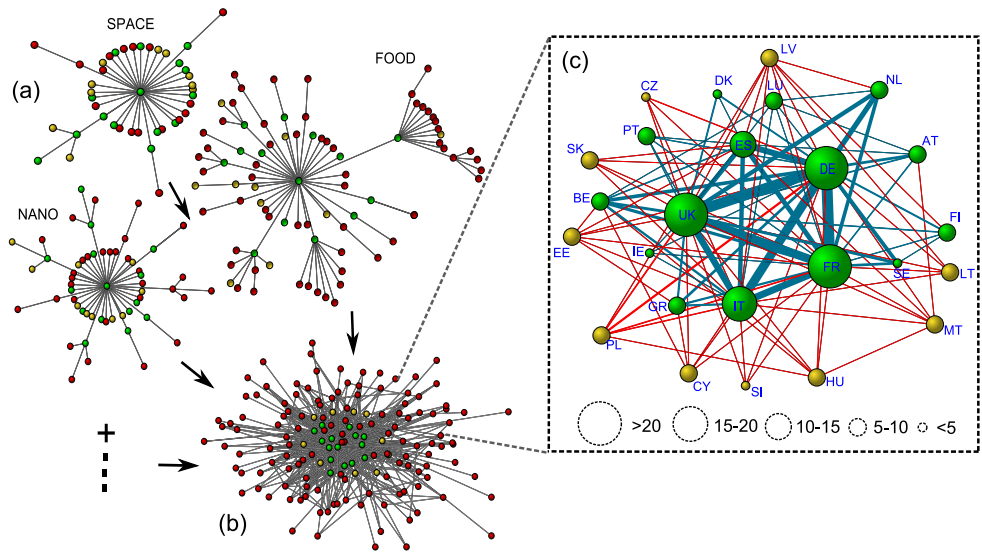}
}
\caption{(color online) (a) Example of some Minimum Spanning Trees (MSTs) created from the collaboration network of the FP6 for different instruments. (b) The Overlapping Tree Network (OTN) obtained from merging all the MSTs for all the 16 separate activities. (c) The OTN that shows only the collaboration activity between the EU25 countries for different instruments of the FP6. With green nodes we represent the EU15 countries and with yellow nodes we represent the 10 new member countries of EU, and with red nodes the countries outside the EU. The sizes of the nodes are proportional to degree $k$ of the node. With blue lines we represent the links between the EU15 countries, and with red lines we represent the links that connect the 10 new EU members to the network. The thickness of the links are proportional to the weight of the link.}
\label{Fig:2}
\end{figure*}

In the MSTs that we calculate, each link has a weight, which is representative of the strength of the interaction between nodes $i$ and $j$. We use these weights of the individual MSTs to calculate the new weights of the combined graph $G$. We set the weight of a link between nodes $i$ and $j$ in graph $G$ as the mean value of the strength that the same link has for the entire set of the $M$ MSTs. Because of this construction method, we name the resulting graph  ``Overlapping Tree Network'' (OTN). A representative example of the above procedure is given in Figure~\ref{Fig:1}. In the following section, we will implement this technique to the different MSTs that we obtain from the European collaboration network of the FP6, and to the network of the 100 highest capitalized stocks traded in the NYSE from 1995 to 2003. 

\section{Application to European collaboration network}

Joint scientific research in Europe has been funded through large programs called Framework Programs (FP). In all FPs partners come from different countries, and thus international collaborations are strongly encouraged. In the following we construct the collaboration networks of all countries participating in the FP6, which is the last concluded FP that ran during the period 2002-2006. Such dataset can be obtained from CORDIS~\cite{bb:CORDIS}. The collaboration networks are constructed separately for every thematic area, out of a total of 16, by considering each country as a node, and by representing the collaborations among countries as links between the nodes. This means that an edge connecting two nodes, $i$ and $j$, represents the presence of at least one collaboration between institutions from country $i$ with institutions from country $j$. 

At each edge we assign a weight $w_{ij}$, that represents the total number of collaborations between institutions of the two countries, on a specific thematic area. We transform this weight to a distance measure $d_{ij}=1/w_{ij}$, in such a way that the smaller the distance, the stronger the collaboration between countries. By default $d_{ij}$ is defined in the interval $\left(0,1\right]$, and takes its maximum value when there is only one collaboration between a pair of countries, $w_{ij}=1$. 

The use of Spanning Trees as subnetworks that retain only the most meaningful connections of the original network, is an approach that has enhanced our understanding in various complex systems. Following this approach in a previous work~\cite{bb:GarasFP} we used the MST to measure the role of a country in the collaboration network that resulted due to its participation in FP6.  We found that each MST of the FP6 collaboration network has star like structure around some specific countries, for different thematic areas. These countries, that are found to act as hubs (strongly connected nodes), are Germany, United Kingdom, France, and Italy. More specifically, we find that Germany (DE) is the central hub for 62.5\% of the thematic areas, United Kingdom (UK) 25\%, France (FR) 6.25\%, and Italy (IT) 6.25\%. 

Here, for each one of the 16 thematic collaboration networks we calculated again these Minimum Spanning Trees, and we combined them applying the aforementioned technique to obtain the OTN of the collaboration network. Examples of MST, where the star like structure becomes apparent and the resulting OTN are shown in Figure~\ref{Fig:2}. The OTN is still a well connected network, but its connections represent only the stronger collaboration links, as they are captured by the MTS for every thematic area. In order to get a more detailed view, we zoom more to the OTN, by examining only the collaborations among the 25 EU member countries. A strongly interconnected community of the most frequent hubs of the network is  identified. This community is like a nucleus of countries with very strong collaboration links, while the other countries play only a satellite role around it. 

This picture is much richer in comparison to the MST because, not only it highlights the importance of the hubs of the network (this information we can get from the MST as well~\cite{bb:GarasFP}), but it also shows quantitatively the interconnecivity pattern between them, in such a way that a fully connected community is formed between DE, UK, FR, IT, and ES.  This means that these five countries not only play important role in the network, but they are connected among themselves with very strong connections, as it is shown by the thickness of the links between them. This valuable information is not possible to be extracted from the MST since, by definition, the MST does not allow loops to be formed. 

\section{Application to correlation based networks}

We now apply the OTN technique to the network of the 100 most capitalized stocks traded in the NYSE, using daily returns in the period 1995 - 2003. The basis of the creation of an equity network, using a portfolio of $N$ stocks, is the analysis of the cross-correlation among time series of returns for all pairs of stocks. The correlation coefficient provides a similarity measure that can be used as weight for a link between each pair of stocks. Thus, a correlation based network is a fully connected weighted graph, and the weights of the graph are obtained from the correlation matrix of the system. The extraction of the MST from such graphs gives a wealth of useful information~\cite{bb:Mantegna,bb:Bonanno,bb:Onnela,bb:Garas,bb:TumminelloChaos}, but because the MST is a very drastic filtering method, it cannot capture more structured entities, such as the communities of stocks connected with strong links~\cite{bb:GarasArgyrakisHavlin}. The need to find filtering techniques that will create richer graphs than the MST was first addressed by Tumminello et al.\ with the creation of the Planar Maximally Filtered Graph (PMFG) technique~\cite{bb:Tumminello}. The PMFG is a graph that is created by adding links to a disconnected network in decreasing weight order, with the constrain that the resulting graph will be planar. In what follows, we will use the OTN technique and we will compare our results with the results of the PMFG.

In order to apply the OTN technique for the case of financial correlation based networks that originate from time series data, as explained in the ``Methods'' section, we divide the time series to a set of smaller segments using a sliding time window of length $T$. The length $T$ of the time window is a fraction $q=T/T_{0}$ of the original size of the return time series, for which $T_{0}=2262$ days. The time step that we use to move the time window is one day. For every time step we calculate the correlation matrix, which contains $N(N - 1)/2$ entries, determined from $N$ time series of length $T$. If $T$ is not very large compared to $N$, it is shown using arguments from Random Matrix Theory~\cite{bb:Laloux-RMT,bb:Plerou-RMT} that the determination of the correlations is noisy. As a results, in order to have a more reliable determination of the correlation matrix, we make sure that $Q=T/N \geq 1$.

\begin{figure}
\resizebox{1\columnwidth}{!}{%
\includegraphics{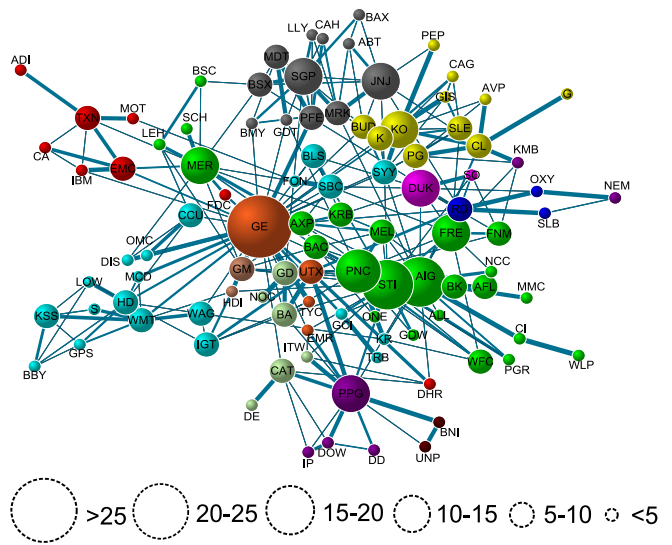}
}
\caption{(color online) OTN obtained from the network of the 100 most capitalized stocks traded in the NYSE in the period 1995 - 2003. For the calculation a sliding time window of 1200 days was used. The different colors represent stocks belonging to different sectors of economic activity, according to the Standard Industrial Classification (SIC) codes. The size of each node is proportional to the degree $k$ of the node. The $k$ values of the nodes are inside the interval $k\in[1,32]$. The thickness of each link is proportional to the weight of the link.}
\label{Fig:3}
\end{figure}

Using this procedure we calculate the OTNs for different time window lengths. As an example we show in Fig.~\ref{Fig:3} the OTN that was calculated using time window of length $T=1200$ days. From this figure we see that most of the stocks belonging to the same sector of economic activity are, as expected, clustered together. But from the OTN of Fig.~\ref{Fig:3} we  identify how communities of stocks belonging to the same sector of economic activity are connected, and how strong the links between the communities are, just by examining the thickness of these links. Furthermore, it is now possible to identify certain stocks that have large number of connections with stocks from different sectors. Such stocks are the stock of General Electric (GE), which is the most connected stock of the network, the stock of American International Group (AIG) that is an insurance and financial conglomerate company, etc.

This shows that these stocks have activities not only in their main field of operations but they also extend to totaly different fields, as for example GE, which is known for electrical machinery (from airplane engines to light bulbs) extending to the financial sector with GE Capital. Our method is able to extract interdisciplinary activities that other methods cannot accomplish effectively. 

\begin{figure}
\resizebox{1\columnwidth}{!}{%
\includegraphics{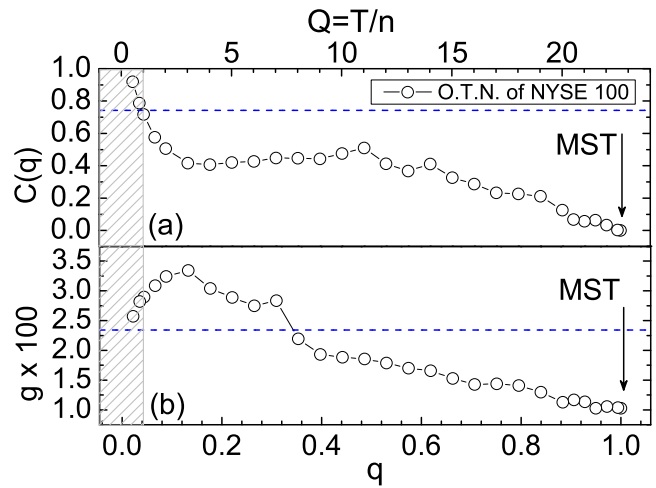}
}
\caption{(a) Clustering coefficient $C$ of the OTN obtained from the network of the 100 most capitalized stocks traded in the NYSE using different sliding time windows versus the fraction $q$ of the time series length that falls inside the time window. (b) Structure measure $g$ of the OTN obtained from the network of the 100 most capitalized stocks traded in the NYSE using different sliding time windows versus the fraction $q$ of the time series length that falls inside the time window. In both plots the horizontal line stands for the respective values of the PMFG obtained from the same data set. The top axis show the values of $Q$, and the shaded area shows the area $Q \leq 1$ where the correlation matrix of the system is dominated by noise.}
\label{Fig:4}
\end{figure}

A widely used technique that quantifies the tendency of the nodes of a graph to cluster is the clustering coefficient $C(q)$ \cite{bb:WattsStrogatz,bb:AlbertBarabasi}, which is defined as follows. Assuming that a vertex $i$ has $n_{i}$ neighbors then at most $n_{i}(n_{i}-1)/2$ edges can exist between all the neighbors of vertex $i$. If we denote with $C_{i}(q)$ the fraction of such existing edges for node $i$, then $C(q)$ is defined as the average of $C_{i}(q)$ over all connected nodes of the network. 
In Fig.~\ref{Fig:4}(a) we plot the values of the clustering coefficient obtained for OTNs constructed using various time window lengths. As it is expected the clustering coefficient $C(q)=0$ for the case of the MST, since there are no loops in a tree. In the same plot the horizontal discontinuous line, for comparison reasons, shows the value of the clustering coefficient obtained using the PMFG technique. As we see from Fig.~\ref{Fig:4}(a) the PMFG, constructed using the same time series length that we used to calculate the MST, has always higher clustering coefficient in comparison to the OTN.

The above result is expected, since PMFG is constructed in such way that many small cliques are formed. But all these cliques are not necessary meaningful, in the sense that a cluster of stocks belonging to the same sector is, since the formation of cliques is forced by the construction algorithm. On the other hand, clustering information is important, because it plays central role in understanding the hierarchical structure of an equity network~\cite{bb:Mantegna}, and could be effectively applied in portfolio selection~\cite{bb:Onnela}. 

In what follows we introduce an empirical function that can be used to measure ho meaningful are the clusters that we get by filtering methods applied to correlation based networks, if we assume that the best method will cluster all stocks of the same sector together. Of course, as we discussed above, there are stocks that are strongly connected to stocks of different sectors of economic activity, but such stocks are usually the exception, not the rule.

In order to calculate the structure of the network we introduce a structure function $g$. This function can take values in the range $[1,0)$, and it obtains the maximum value $g=1$ for a complete graph that all its nodes belong to the same partition. The $g$ function is defined as,
\begin{equation}
	g=\frac{1}{N_{L} N_{n}}\sum_{j=1}^{s}\frac{1}{N_{j}N_{cl}^{j}N_{L}^{j}}\sum_{i}^{N_{cl}^{j}}\left(m_{cl}^{i}l_{cl}^{i} \right)^2
\label{eq:g1}
\end{equation}
where $N_{L}$ is the total number of links of the network, $N_{n}$ is the total number of nodes of the network, $s$ is the number of different partitions (in this case is the number of different economic sectors), $N_{j}$ is the number of nodes that belong to partition $j$, $N_{cl}^{j}$ is the number of different clusters that are created from nodes belonging to partition $j$, $N_{L}^{j}=N_{j}(N_{j}-1)/2$ is the number of possible links between all the nodes of partition $j$, $m_{cl}$ is the number of nodes that form cluster $cl$, and $l_{cl}$ is the number of links that connect the nodes of cluster $cl$. We should make clear that in order to use the structure function $g$, the partitioning of the network must be known {\it a priori}, therefore this function does not give information about the partitioning of a graph, it only compares the output of different partitioning methods.

We applied this measure to extract the structural information obtained using the OTN for different time windows and we compared it with the structural information obtained using the PMFG. The results are shown in Fig.~\ref{Fig:4}(b), where we see that for large time window lengths the OTN is lower in comparison to the PMFG, but by decreasing the time window length eventually the OTN is able to capture more structural properties than the PMFG. This happens because the OTN is based on the MST, which is a relatively stable structure. Thus, for large time window lengths the OTN includes only a small fraction of the links of the original fully connected graph, just a few more than the links included in the MST. But as we decrease the time window length, we capture more information about the evolution of links that become strong only for some period of time. Such links are included in the OTN and make it optimal for studying the network at different time periods. If we compare the set of the common links between the MST that is calculated using the full length time series, and the OTNs that we have calculated for all the time window lengths that are shown in Fig.~\ref{Fig:4}, we find that on average $98.7\pm0.2$\% of the links included in the MST is also included in the OTN. This means that almost all the connectivity information that we extract from the MST is included in the OTN, but the OTN allows to examine even more details of the system.

\section{Discussion}

We introduced a general method that extracts information stored in complex networks, by using only a subset of the network strongest links, resulting in the Overlapping Tree Network (OTN). Our method is based upon the well established technique of the extraction of the Minimum Spanning Tree, but it allows for the filtered network to have loops, and therefore it retains more of its original complexity that the MST. The added information in the OTN is that besides the clustering together of similar nodes as in the MST, we now have the full picture of the connectivity pattern between the hub nodes, and the strength of these connections ($w_{ij}$). Such strongly interconnected clusters form easily distinguishable communities that can be used to partition the network. Furthermore, the OTN includes links that could be strong for only a certain time period or for some specific set of interactions, and then become weak again. Such dynamical transition is not detected by the use of the MST alone. As a consequence the importance of some central nodes of the network detected by the MST, such as, for example the stock of GE, is highlighted by the use of OTN. This method was applied to two different systems and gave interesting insights for both cases.

The first system was the collaboration network of countries participating to at least one of European sponsored research projects. With this approach we were able to identify a fully connected community of the most frequent hubs of the network. The strength of the internal links connecting the nodes of this community are much higher than the average strength of the links of the remaining network. Therefore this structure could be described as a nucleus of countries with very strong collaboration links. All other participating countries were found to play a satellite role around this nucleus. Such information is not palpable by the use of the MST alone because with the MST approach we lose information about the interconnectivity. But interconnectivity information is very important both for policy makers and for scientists that create consortia and apply for funding to the European Commission.

The second system was the network of the 100 most capitalized stocks traded in the NYSE. These stocks form a correlation based network, which is calculated using time series of the daily returns of these equities. The OTN was extracted using different lengths of the sliding time window. We verified the expected clustering of stocks according to the sectors of economic activity that they belong, but we were able to identify the stocks of General Electric (GE), and the American International Group (AIG) as the ones that have the strongest links with stocks from different sectors in the set of 100 highest capitalized stocks. Since clustering is an important information, we introduced an empirical function that is able to quantify the result, in terms of clustering information, of different filtering techniques applied in correlation based networks. We applied this function to the outcome of the OTN and of the PMFG methods. Comparison of these methods showed that the PMFG gives better clustering information if the length of the time window used for the OTN method is large. For smaller time windows the OTN is able to capture more structural properties, and it is more suitable for studying the network at different time periods.

\section{Acknowledgments}
The authors would like to thank M. Tumminello for his valuable comments during the preparation of this work, and L.K. Gallos for his contribution in the preparation of the EU database. 
This work was partially supported by a European research NEST/PATHFINDER project, DYSONET 012911, and by the Greek General Secretariat for Research and Technology of the Ministry of Development, PENED project 03ED840.


\end{document}